\documentclass[%
reprint,
 amsmath,amssymb,
 aps,
floatfix,
]{revtex4-2}

\usepackage{graphicx}
\usepackage{dcolumn}
\usepackage{bm}
\usepackage{xcolor}
\usepackage{siunitx}
\usepackage{comment}
\usepackage[normalem]{ulem}

\begin{document}

\preprint{APS/123-QED}

\title{Unlocking optical coupling tunability in epsilon-near-zero metamaterials through liquid crystal nanocavities}



\author{Giuseppe Emanuele Lio$^{1,2,\circledast}$}
 \email{lio@lens.unifi.it}
\author{Antonio Ferraro$^{3\circledast}$}%
\author{Bruno Zappone$^3$}
 \email{bruno.zappone@cnr.it}
\author{Janusz Parka$^4$}
\author{Ewa Schab-Balcerzak$^5$}
\author{Cesare Paolo Umeton$^{3,6}$}
\author{Francesco Riboli$^{2,7}$}
\author{Rafa{\l} Kowerdziej$^4$}
\author{Roberto Caputo$^{3,6,8}$}
\email{roberto.caputo@unical.it}

\affiliation{%
$^1$ Physics Department, University of Florence, 50019, Sesto Fiorentino, Florence, Italy\\
$^2$ European Laboratory for non Linear Spectroscopy (LENS), 50019, Sesto Fiorentino, Florence, Italy\\
$^3$ Consiglio Nazionale delle Ricerche - Istituto di Nanotecnologia (CNR-Nanotec), Rende (CS), 87036 Italy \\
$^4$ Institute of Applied Physics, Military University of Technology, 2 Kaliskiego Str., 00-908, Warsaw, Poland \\
$^5$ Centre of Polymer and Carbon Materials Polish Academy of Sciences, 34 M. Curie-Sklodowska Str., 41-819 Zabrze, Poland \\
$^6$Physics Department, University of Calabria, \\87036 Arcavacata di Rende (CS), Italy \\
$^7$ National Institute of Optics, CNR-INO, 50019, Sesto Fiorentino (FI), Italy\\
$^8$ Institute of Fundamental and Frontier Sciences, University of Electronic Science and Technology of China, Chengdu 610054, China\\
$\circledast$ Authors contribute equally to this work. \\
}%

\begin{abstract}
Epsilon-near-zero (ENZ) metamaterials represent a powerful toolkit for selectively transmitting and localizing light through cavity resonances, enabling the study of mesoscopic phenomena and facilitating the design of photonic devices. In this experimental study, we demonstrate the feasibility of engineering and actively controlling cavity modes, as well as tuning their mutual coupling, in an ENZ multilayer structure. Specifically, by employing a high-birefringence liquid crystal film as a tunable nanocavity, the polarization-dependent coupling of resonant modes with narrow spectral width and spatial extent was achieved. Surface forces apparatus (SFA) allowed us to continuously and precisely control the thickness of the liquid crystal film contained between the nanocavities and thus vary the detuning between the cavity modes. Hence, we were able to manipulate nanocavities anti-crossing behaviors. The suggested methodology  unlocks the full potential of tunable optical coupling in epsilon-near-zero metamaterials and provides a versatile approach to the creation of tunable photonic devices, including bio-photonic sensors, and/or tunable planar metamaterials for on-chip spectrometers.
\end{abstract}

\maketitle

\section{Introduction}
In the eighteenth century, the voltaic pile invented by Alessandro Volta demonstrated that stacking materials with different properties can lead to groundbreaking devices with significantly novel functionalities. Nowadays, this approach is recognized as a cornerstone of fabrication technology, particularly in the development of  high-performance nano-devices. The Fabry-Perot resonator is one of the most convenient and broadly used devices in photonics, particularly for engineering light-matter coupling \cite{reveret2008influence, Lio2019cavities, cao2021strong} and is commonly used in color filters \cite{choudhury2016experimental, lio2020color}, two-photon direct laser writing with hyper-resolution \cite{guo2018focusing, lio2021leveraging}, optical metasurfaces \cite{shaltout2016development, kowerdziej2022soft}, high-heat release \cite{dyachenko2016controlling, ferraro2021tailoring}, sensing devices \cite{sreekanth2016extreme, sreekanth2013sensitivity, lio2023engineering}, and anti-counterfeiting tags \cite{ferraro2022hybrid}, just to name a few. The resonant cavity is usually fabricated by sandwiching a transparent dielectric layer between two partially reflecting mirrors. These metal-dieletric resonators possess the intriguing properties of epsilon-near-zero (ENZ) effective permittivity \cite{vassant2012berreman, reshef2019nonlinear, wu2021epsilon} at specific resonance wavelengths that can be finely tuned by carefully selecting the thickness and refractive index of the metal and dielectric layers, and the angle and polarization of the incoming light  \cite{Lio2019cavities}. Multilayer resonators also allow to efficiently manipulate electromagnetic waves in specific spectral ranges and enable optimal solutions for device miniaturization \cite{bilotti2010}, fabrication of perfect absorbers for structural coloring in the VIS-NIR range, \cite{li2015}, and high photovoltaic conversion \cite{heydari2017}. Furthermore, photon confinement in optical nanocavities enables an efficient control of light-matter coupling in fundamental physics studies of single quantum objects \cite{imamog1999quantum} and correlated polaritons \cite{greentree2006quantum, patra2023plane}, as well as applications in quantum optical devices, and sensors \cite{vahala2003optical,liang2013scalable,frisk2019ultrastrong,herzog2020realization,smith2020active}.

In this context, there is high demand of devices that can be reconfigured and adapted to various emerging technologies, especially in the automotive and telecommunication sectors \cite{cui2019tunable}. Current ENZ metamaterial technologies, however, lack the ability to dynamically adjust their functionalities. Liquid crystals (LCs) show a large and fast response to external stimuli and are ideal candidates to overcome this limitation. For instance, elastomeric LCs have been used to tune photonic crystals \cite{de2023temperature} and Fabry-Perot cavities \cite{zubritskaya2023dynamically}. LC-based metasurfaces have also been recently implemented, confirming the extraordinary capabilities of LCs in the active control of visible light \cite{sharma2020all, Lio_photonics8030065, wang2022metasurface, palermo2022all}, while extensions to the microwave \cite{kowerdziej2013,zografopoulos2019liquid} and terahertz regimes \cite{kowerdziej2015,isic2019electrically} are under way. The primary challenge to developing an active, LC-based ENZ metamaterial is to reduce the LC thickness to a few hundred nanometers. This thickness is considerably smaller than the limit of a few micrometers currently achieved in display technology.\\
In this article, we present experimental and numerical evidence of optical coupling in ENZ multilayer metamaterials comprising a nanoscale high-birefringence LC film with tunable thickness achieved by means of a Surface Force Apparatus (SFA). Originally designed to measure surface forces across fluid films \cite{israelachvili_mcguiggan_1990}, the SFA has been recently introduced in photonics as a tool to control mode coupling in optical cavities \cite{Zappone_ACSPhotonics_2021,patra2023plane}. Specifically, we have investigated a system comprising a nanoscale LC film (T-layer) with variable thickness $d_T$ sandwiched between two identical metal-insulator-metal (MIM) cavities, thereby creating a symmetric three-cavity resonator denoted as MIMTMIM. The MIM cavities were fabricated by sputtering deposition on two cylindrical surfaces having a radius $R = 2$ cm. The surfaces were mounted with crossed axes in the SFA ensuring a single contact point (i.e., point of closest surface approach, $r = 0$ in Fig. \ref{1}b) where the surface distance was $d_T$. Around this point, the distance $h_T$ varied approximately as in a sphere-plane geometry: $h_T \approx d_T + r^2/2R$. The three-cavity resonator was illuminated with white light under normal incidence. The SFA allowed controlling the LC thickness dynamically, accurately, and continuously from several tens of microns down to the direct mechanical contact between the MIM surfaces (Fig. \ref{1}a). Details about the SFA technique are provided in the \textit{Materials and method} section and a scheme is shown as supporting information (Fig. \ref{SFA_setup}).

\section{Mode coupling in a multi-cavity resonator}
\begin{figure}[!ht]
\includegraphics[width=0.9\columnwidth]{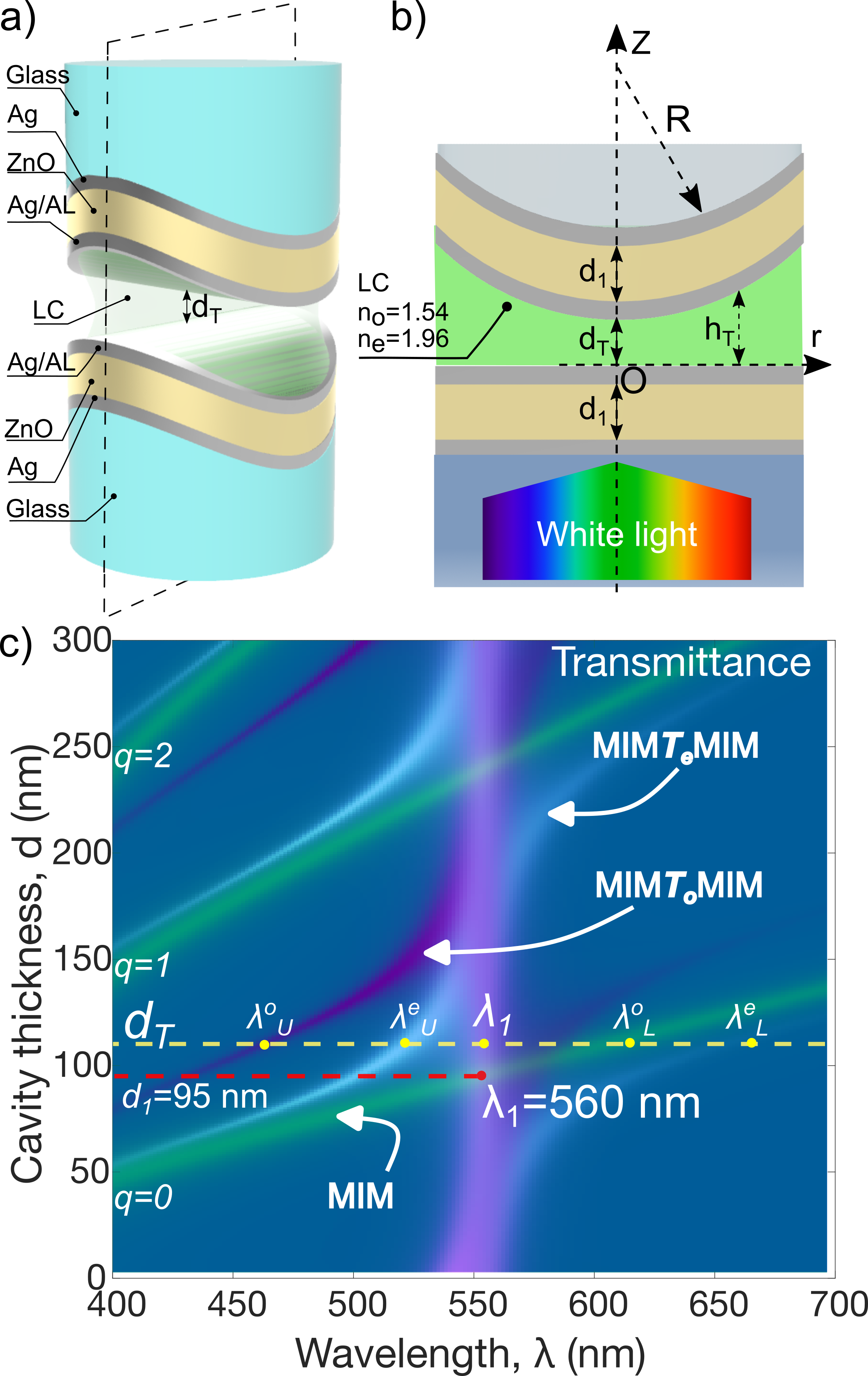}
\caption{Illustration of the three-cavity MIMTMIM resonator realized in the SFA. A ZnO layer (I) sandwiched between two Ag layers (M) with equal thickness constitutes a MIM cavity and was deposited by sputtering on each of the two cylindrical glass lenses of the SFA. LC film (T) confined between the MIM cavities forms a third cavity with non-uniform thickness. b) Cross-section view of the SFA geometry with the $z$ direction of light incidence is evidenced. The geometry of the two cylindrical surfaces is approximated by a plane and a sphere with $r$ being the lateral distance from the contact point ($r=0$). $h_T$ and $d_T$ are respectively the surface distances at a given $r$ and at $r=0$. c) Transmittance under normal incidence calculated using the TMM method for a MIM cavity (green lines) as as function of the wavelength $\lambda$ and thickness of the I-layer, and for the three-cavity resonator as a function of $\lambda$ and thickness $d_T$ of the LC film (T-layer), for both ordinary (MIMT$_o$MIM, purple curves) and  extraordinary polarizations (MIMT$_e$MIM, cyan curves). A resonance corresponds to a local intensity maximum. $\lambda_1$ is the first-mode wavelength of the MIM cavity obtained for the thickness $d_1$. In addition to $\lambda_1$, four different wavelengths $\lambda^{o,e}_{U,L}$ are obtained at the LC thickness $d_T$ marked with a yellow horizontal line, where $o$ and $e$ indicate the given polarization and $U$ and $L$ denote the (lower/upper) photon energy relative to $\lambda_1$.}
\label{1}
\end{figure}

Let us begin the theoretical considerations with the analysis of multi-beam interference under normal incidence in a single (Fabry-Perot) MIM cavity constituted by an isotropic material (I-layer). A plane wave resonates with a cavity if the following condition of constructive interference occurs \cite{fowles1989modern, born2013principles}: 
\begin{equation}
 n_T K_{q}d_T=q\pi-\phi
\label{eq:2}
\end{equation}
where $n_T$ is the refractive index of the cavity medium, $d_T$ is the metal-metal surface separation distance, $q$ is the resonance order, $K_q$ = $2\pi/\lambda_q$ is resonance wavevector with wavelength $\lambda_q$, and $\phi$ is the phase shift due to reflection at the dielectric-metal interface. Both $n$ and $\phi$ vary slowly with the wavelength and can be considered approximately constant across the $\sim$ 100 nm spectral range of an SFA experiment. The resonance condition Eq.\ref{eq:2} can thus be rewritten as:
\begin{equation}
 \lambda_q=2n_Td_T/(q-\phi/\pi)
\label{eq:3}
\end{equation}
showing that the resonance wavelength $\lambda_q$ increases linearly as the surface distance $d_T$ or the refractive index $n_T$ increases, whereas it decreases when the order number $q$ increases. The transmittance of a MIM cavity under normal incidence can be accurately calculated as a function of wavelength $\lambda$ and the cavity thickness using the transfer matrix multiplication (TMM) method (green lines in Fig. \ref{1}b).  For the MIM cavities considered in our experiments, only one resonance wavelength $\lambda_1 = \SI{560}{\nano \meter}$ appeared in the SFA spectral range (vertical dashed black line in Fig. \ref{1}b). The TMM calculation showed that $\lambda_1$ corresponded to the first resonant mode obtained for the MIM cavity thickness of $d_1$= \SI{95}{\nano \meter} (horizontal dashed black line in Fig. \ref{1}b). 

In a three-cavity MIMTMIM resonator, a cavity mode can overlap and interfere with the resonances of neighboring cavities across the metal (M) layers. Consequently, the coupling of resonances and the optical interaction among the cavities that give rise to hybrid resonance modes \cite{Zappone_ACSPhotonics_2021}. In the MIMTMIM resonator, hybridization is strongest between the outer MIM cavities and the central MTM cavity when the resonance wavelength $\lambda_1$ of the former crosses the resonance wavelength $\lambda_q$ of the latter. If the central liquid crystal cavity, T-layer indicated as ($d_T$), is optically isotropic, hybridization creates a threefold splitting of the resonance, i.e., a triplet of wavelengths $\lambda_L, \lambda_1$, and $\lambda_U$ with increasing photon energy \cite{Zappone_ACSPhotonics_2021}. On the other hand, when the T-layer is an optically anisotropic LC film, these wavelengths depend on light polarization (purple and cyan curves in Fig. \ref{1}b). In contrast with Eq. \ref{eq:2}, the resonance wavelengths $\lambda_L$ and $\lambda_U$ do not vary linearly with the thickness $d_T$ in proximity of the crossing point between $\lambda_1$ and $\lambda_q$. However, as the wavelength $\lambda_q$ departs from $\lambda_1$, hybridization weakens and $\lambda_L$ and $\lambda_U$ converge towards $\lambda_q$ thus acquiring an almost linear dependence on $d_T$. These preliminary considerations highlight the potential of birefringent materials for tuning the resonances of metal-dielectric metametarials. 

\begin{figure}[!htb]
\includegraphics[width=0.9\columnwidth]{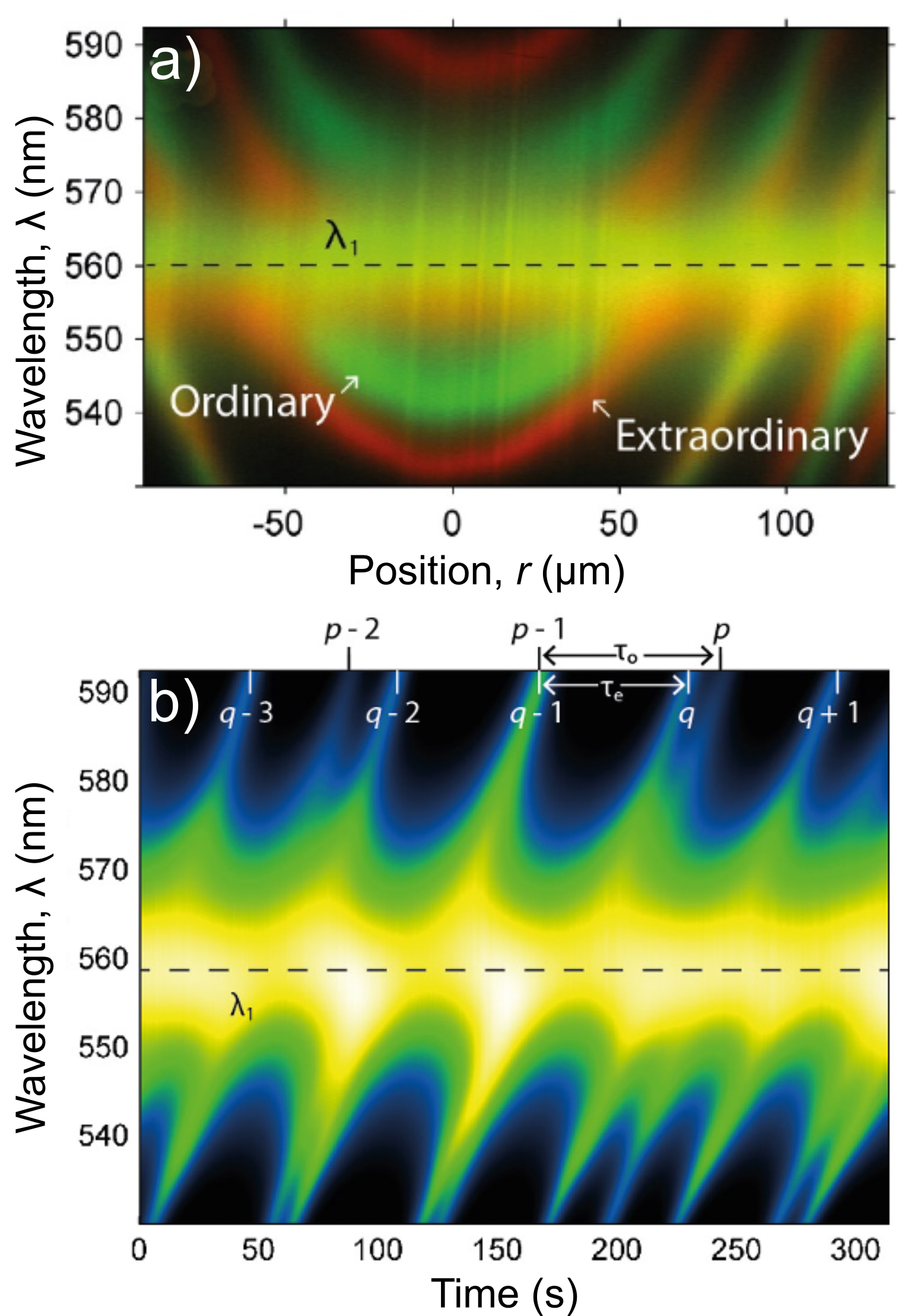}
\caption{a) Transmitted intensity $T$ measured for the MIMTMIM resonator in the SFA as a function of the wavelength $\lambda$ and lateral distance $r$ from the surface contact position ($r=0$). Green and red fringes correspond to ordinary and extraordinary polarization, respectively, and were obtained by setting a linear polarizer perpendicular or parallel to the planar LC anchoring direction. The two fringe types overlap in the yellow regions, corresponding to the first-order resonance wavelength $\lambda_1$ of the MIM cavities. b) Transmitted intensity measured at the contact position ($r = 0$) as a function of time $t$ and wavelength $\lambda$ while separating the surfaces at constant speed $u$. Each value of $t$ corresponds to a different separation distance (i.e., LC film thickness): $d_T= d_{0} + ut$, where $d_0$ is the initial surface separation and $u$ is the (constant) speed of surface separation. The mode order for ordinary and extraordinary fringes is denoted as $q$ and $p$, respectively. A fringe with order $q$ (or $p$) exits the spectral range at wavelength $\lambda$ = \SI{593.2}{\nano \meter} after a time $\tau$ compared to the fringe with order $q\:-\:1$ (or $p\:-\:1$). The delay is $\tau_o\:=\:78.35$ s and $\tau_e\:=\:60.85$ s for ordinary and extraordinary fringes, respectively.}
\label{5}
\end{figure} 

\section{Experimental results}

In the realized system, the thickness of silver (Ag, M-layers) and zinc oxide (ZnO, I-layers) is \SI{30}{\nano \meter} and \SI{95}{\nano \meter}, respectively. The LC material considered for the T-layer is a high-birefringence nematic liquid crystal mixture named LC1825, synthesized by the Military University of Warsaw \cite{dabrowski2013high} with a birefringence of $\Delta n= n_e - n_o = 0.42$, where $n_e=1.96$ and $n_o=1.54$ are the extraordinary and ordinary refractive indices at room temperature, respectively. The photoalignment compound JK158 \cite{wkeglowski2015poly} was spin-coated on the Ag surfaces facing the LC to induce planar orientation along the cylinder axis on one surface and perpendicular to the axis on the other surface. Crossing the axes in the SFA ensured a planar alignment uniform across the LC thickness. Therefore, polarized parallel or perpendicular to the LC orientation travelled in the LC film as purely extraordinary or ordinary waves, respectively. Further details on the fabrication and materials used in our experiments are provided in the \textit{Materials and methods} section. 
During the experiment, a collimated white-light beam coming from a halogen lamp illuminated the MIMTMIM resonator under normal incidence and the transmitted intensity was analyzed using an imaging spectrograph coupled to a high-resolution CCD camera. In the spectrogram of  Fig. \ref{5}, the transmitted intensity $I$ was measured as a 2D function of the wavelength $\lambda$ and lateral distance $r$ for the contact point ($r=0$ in FIg. \ref{1}). A resonance produces a local maximum in the intensity function $I_0(r,\lambda)$. Because $h_T \approx d_T + r^2/2R$ around the contact point, resonance wavelengths that depend linearly on $h_T$ vary quadratically with $r$ and create curved fringes in the SFA spectrograms with a parabolic tip corresponding to the contact point. 
In the spectrogram of Fig. \ref{5}b, the intensity was measured at the contact position ($r=0$) while increasing the surface distance $d_T$ at a constant speed $u$ of a few nm/s using a motorized actuator (Fig. \ref{1}). In this case, the intensity $I_0$ was resolved as a 2D-function of $\lambda$ and time. Because $d_T(t) = d_0 + ut$,where $d_0$ is the initial thickness, each vertical line in the spectrogram corresponds to a specific time $t$ and surface distance $d_T(t)$.  The advantage of this approach is that the SFA can vary $d_T$ dynamically and continuously over a wide range of surface distances, from several $\mu$m to direct surface contact ($d_T < 1$ nm for molecularly smooth surfaces), with an accuracy better than 1 nm and execution time of the order of minutes. To vary $d_T$, the surfaces were approached to or separated from each other at a constant speed. By recording $I_0(\lambda,t)$, the SFA allowed studying the resonance dispersion as a function of the cavity thickness $d_T$ in a single sweep of thickness, instead of fabricating multiple cavities with different thicknesses.
In the spectrograms of Fig. \ref{5}a,b, the first-order resonance of the fixed-thickness MIM cavities produces a specific resonance wavelength $\lambda_1$ that does not depend on the thickness ($d_T$ or $h_T$) of the MTM cavity, i.e., the LC film. On the other hand, the other fringes in the spectrogram  are due to resonances of the MTM cavity and, therefore, depend both on the film thickness and polarization of the incident light. For a fixed surface distance ($d_T$ or $h_T$) and far from $\lambda_1$, these fringes show an approximately parabolic shape (Fig. \ref{5}a) reflecting the surface curvature, as expected. Due to the LC birefringence, these fringes form two distinct sets that can be separately extinguished using a linear polarizer parallel or perpendicular to the planar anchoring direction, as shown in Fig. \ref{5}a (see also an example of unpolarized spectrogram in Fig. \ref{SI_exp} of SI). This finding demonstrates that the resonance modes of the MTM cavity are linearly polarized along the ordinary and extraordinary axis of the uniformly aligned LC. The extraordinary modes appear slightly brighter than the ordinary ones (Fig. \ref{SI_exp} in SI), because light was directed into the spectrograph using a right-angle mirror with polarization-dependent reflectivity (Fig. \ref{SFA_setup} in SI).

\begin{figure}[!ht]
\includegraphics[width=0.5\textwidth]{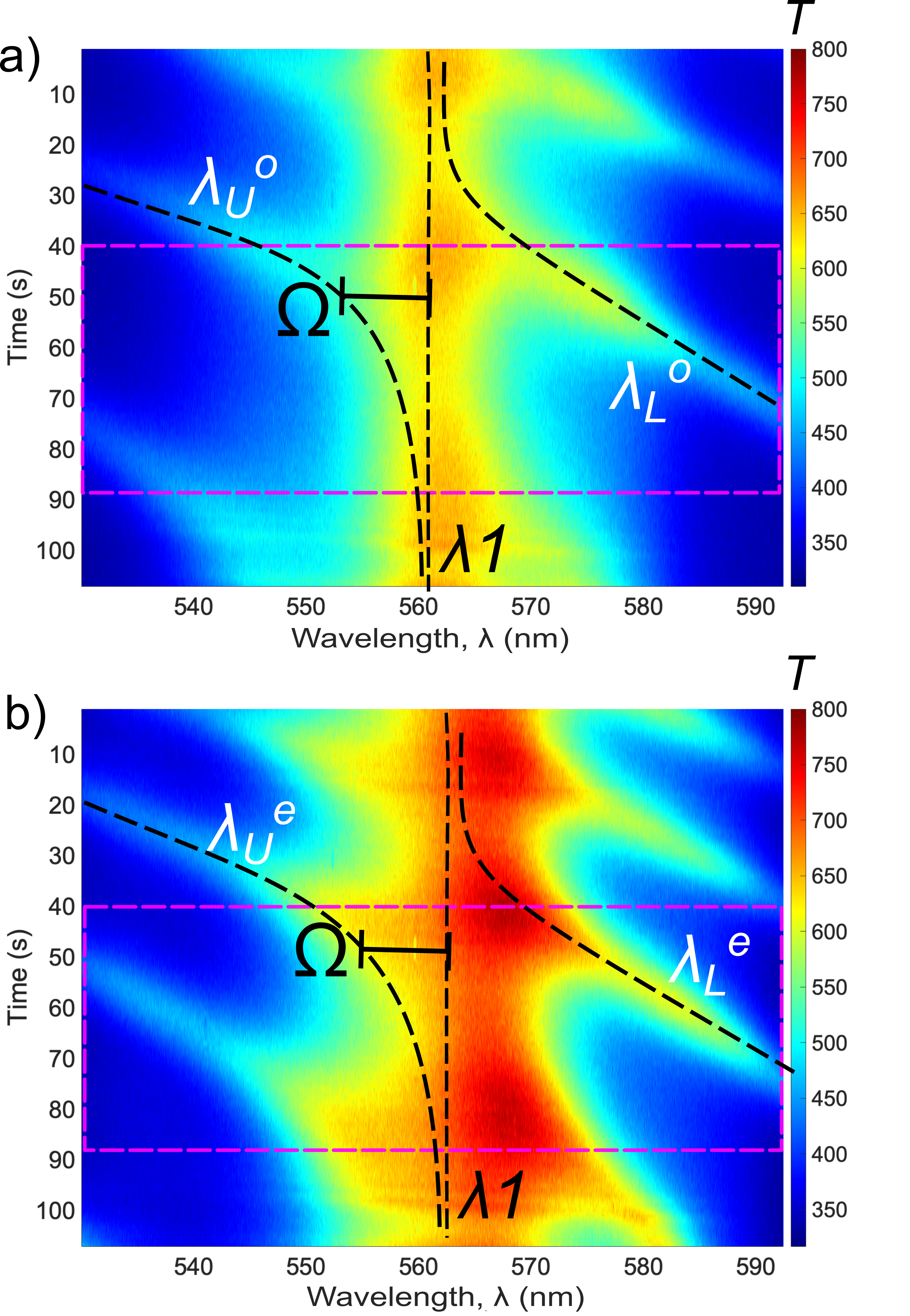}
\caption{Light intensity $T$ transmitted through the system measured for a) ordinary and b) extraordinary polarizations as a function of the wavelength $\lambda$ and time $t$ during surface separation (LC cavity expansion) with constant surface speed $u$. The maps show the resonance wavelengths $\lambda_U$ and $\lambda_L$ obtained during the surface separation together with their separation $\Omega$ from the resonance $\lambda_1$.} 
\label{7}
\end{figure} 

This allows to identify the fringe polarization even though we cannot use a polarizer to dynamically resolve the polarization while varying $d_T$. Figure \ref{5}b shows that extraordinary fringes enter and exit the spectral range of the SFA (at a wavelengths distant from $\lambda_1$) more rapidly than ordinary fringes. In the entrance and exit regions, resonances approximately follow Eq. \ref{eq:2}. Therefore, if the mode with order $q$ resonates at a given wavelength $\lambda_1$, then the mode with order $q \pm r$ resonates at the same wavelength after displacing the surfaces by a distance $\Delta d = \pm r \lambda_1/2n$. As a result of the inequality $n_e > n_o$, extraordinary fringes with index $n_e$ cross the wavelength $\lambda_1$ more often than ordinary fringes as the distance $d_T$ is increased. If the surfaces are separated at a constant speed $u$, the fringe with order $q$ exits the spectral range after a time $\tau =\Delta d/u$ compared to the fringe with order $q - 1$. Figure \ref{5}b shows ordinary fringes exiting the spectral range at wavelength \SI{593.2}{\nano\meter} at periodic time intervals $\tau_o =78.3$ s, whereas the period is $\tau_e =60.8$ s for extraordinary fringes. The ratio of these two periods, $\tau_0/\tau_e=1.29$, is in good agreement with the value $\tau_o/\tau_e\:=\:n_e/n_o\:=\:1.27$ predicted by Eq. \ref{eq:2} using the nominal refractive indices (at room temperature) $n_e\:=\:1.96$ and $n_o\:=\:1.54$ of the LC.

Figures \ref{7}a and \ref{7}b show intensity spectrograms $I_0(\lambda,t)$ obtained for the ordinary and extraordinary polarization, respectively, by using a polarizer in transmission. Resonance wavelengths are highlighted by black dashed lines and are referred to as $\lambda_1$, $\lambda_U$ and $\lambda_L$.
As the thickness $d_T$ of the LC film increases, $\lambda_U$ approaches $\lambda_1$  while $\lambda_L$ departs from $\lambda_1$. Eventually, $\lambda_U$ and $\lambda_L$ become equally spaced from $\lambda_1$ by a distance $\Omega$. This behaviour agrees with the numerical prediction and demonstrate the possibility of dynamically tuning the modes at different wavelengths ranging from \SI{530}{\nano \meter} to \SI{590}{\nano \meter} by acting on LC thickness or incoming light polarization. 
\section{Tuning mode coupling via LC confinement and reorientation}
\begin{figure*}[!ht]
\includegraphics[width=\textwidth]{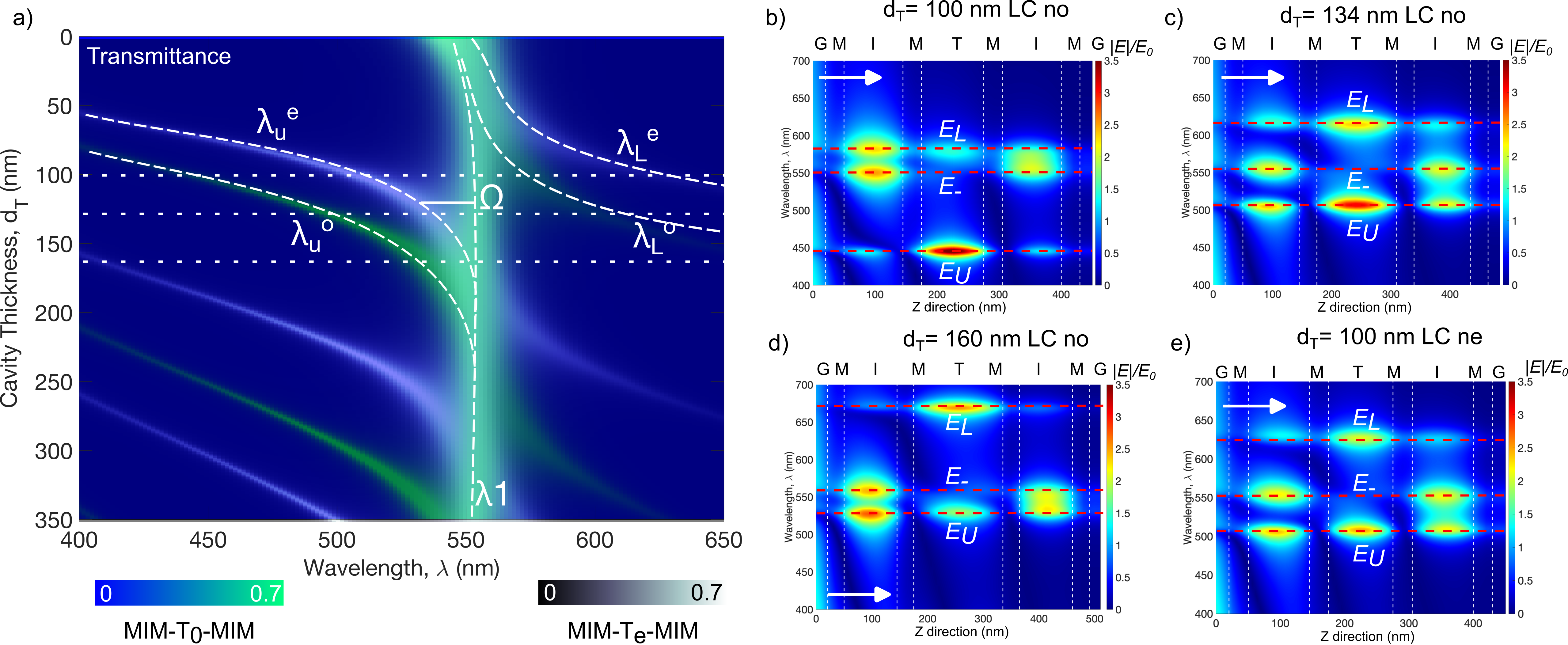}
\caption{a) Calculated spectrogram showing the transmitted intensity as a function of the wavelength $\lambda$ and of the LC cavity thickness $d_T$ for both ordinary (blue-to-green colormap in the bottom left corner) and extraordinary polarizations (black-to-dark cyan colormap in the bottom right corner). b-d) Normalized electric field amplitude ($|E|/E_0$) calculated for the ordinary polarization (LC$_{no}$) as a function of the position $z$ along the surface normal of the MIMTMIM resonator and wavelength $\lambda$. The considered cavity thickness $d_T$ is indicated above each image. e) Electric field calculated for $d_T=100$ nm and extraordinary polarization (LC$_{ne}$). The white dotted vertical lines mark the metal-dielectric interfaces, G (glass) is the medium outside the resonators. The white arrow shows the direction of incidence.}
\label{Sim_no}
\end{figure*}

In order to understand why mode coupling produces a wavelength triplet, we calculated the transmitted intensity as a function of wavelength $\lambda$ and LC film thickness $d_T$ using the TMM method (Fig. \ref{Sim_no}a), and selected three different values of $d_T$ to compute, by a finite element method (COMSOL), the electric field map along the direction perpendicular to the MIMTMIM resonator as a function of $\lambda$ and $z$ position \cite{Lio2019cavities} (Fig. \ref{Sim_no}(b-e)).  

For $d_T=100$ nm and ordinary polarization (Fig. \ref{Sim_no}b), the high-energy mode has wavelength $\lambda_U^o \sim$ \SI{450}{\nano \meter} and is farther away from $\lambda_1$ than the low-energy mode with wavelength $\lambda_L^o \sim$ \SI{585}{\nano \meter}.  This unequal wavelength spacing is reflected in mode hybridization. Namely, the high-energy mode is mainly located in the central MTM cavity while the low-energy mode is more delocalized among the central and outer (MIM) cavities. When the LC film thickness is increased to $d_T=134$ nm (Fig. \ref{Sim_no}c), the high-energy wavelength (Fig. \ref{Sim_no}c) $\lambda_U^o \sim$ \SI{510}{\nano \meter} and low-energy wavelength $\lambda_L^o\sim$ \SI{620}{\nano \meter} are almost equally spaced from $\lambda_1$ and  show a comparable degree of delocalization. When the LC film thickness is further increased to $d_T=160$ nm (Fig. \ref{Sim_no}d), the situation shown in Fig. \ref{Sim_no}b is reversed and the high-energy mode at $\lambda_U^o \sim$ \SI{530}{\nano \meter} is closer to $\lambda_1$ and more delocalized than the low-energy mode at $\lambda_L^o \sim$ \SI{670}{\nano \meter}.
Mode hybridization in MIMTMIM resonator can be understood based on its mirror symmetry under reflection on the middle plane of the central T-cavity. Symmetry requires that resonances be either even (+) or odd (-) under reflection (Fig \ref{Sim_no}(b-e)). Using first-order perturbation theory or variational method \cite{caligiuri2019hybridization,Zappone_ACSPhotonics_2021}, these modes can be approximated as symmetry-adapted linear combinations of single-cavity modes. In particular, the field $E_c$ of first-order mode in the central cavity is even and, therefore, hybridizes with the field $E_+$ of the even combination of outer-cavity modes. Against, the odd combination $E_-$ cannot hybridize with an even mode such as $E_c$. While the modes $E_c$ and $E_+$ overlap and interfere with each other, particularly within the metal layers of the central MTM cavity, direct overlap and interference between the outer cavities is negligible and, as a result, the wavelength $\lambda_-$ of the $E_-$ mode is very close to the wavelength $\lambda_1$ of an isolated MIM cavity. Indeed, the difference between $\lambda_-$ and $\lambda_1$ was too small to be detected in our experiments.

Hybridization between same-parity modes produces the wavelengths $\lambda_L$ and $\lambda_U$ observed both in the SFA experiments and in our calculation \cite{Zappone_ACSPhotonics_2021}. For first-order modes, these wavelengths correspond to the modes $E_L=E_{c}+\alpha E_{+}$ and $E_U=E_{c} - \beta E_{+}$, respectively, where the positive linear coefficients $\alpha$ and $\beta$ depend on the thickness $d_T$ of the central MTM cavity. The wavelengths $\lambda_L$ and $\lambda_U$ are due to the anti-crossing interaction between the even mode $E_{c}$ and $E_{+}$ occurring as $d_T$ varies (Fig \ref{Sim_no}(b-d)). Namely, the $E_U$-mode repels the $E_L$-mode as it moves towards lower energies, while the $E_{-}$ mode is unaffected. The avoided-crossing point is reached when the wavelength $\lambda_q$ of the $E_{c}$-mode (Eq. \ref{eq:2}) overlaps with the first-order wavelength $\lambda_1$ of the outer MIM cavities. In other words, the difference or "detuning" between the photon energies of the two modes becomes zero. At this point, the modes $E_L$ and $E_U$ become uniformly delocalized across the resonator, with equal intensity maxima in each cavity ($\alpha$,$\beta\:\approx\:1$, Fig \ref{Sim_no}c). At the avoided-crossing point, the wavelengths $\lambda_L$ and $\lambda_U$ are found at an equal distance $\Omega$ from the wavelength $\lambda_1$ of the $E_{-}$ mode.

A decisive advantage of using an anisotropic LC film is that the detuning can be actively controlled not only by varying the MTM cavity thickness, but also by selecting the refractive index of the LC. This fact is highlighted in Eq. \ref{eq:2} showing that the resonance wavelength $\lambda_q$ depends on the product $n_Td_T$ and, therefore, the thickness $d_T$ and index $n_T$ play equivalent roles. For example, the transmittance variation obtained by increasing the film thickness from $d_T=$ 100 nm (Fig. \ref{Sim_no}b) to $d_T=$ 134 nm (Fig. \ref{Sim_no}c) can also be obtained by switching the refractive index from ordinary to extraordinary (Fig. \ref{Sim_no}d) while keeping the film thickness fixed to $d_T=$ 100 nm. The switching can be obtained by changing the polarization from ordinary to extraordinary, or acting on the LC orientation (e.g., by applying a voltage to the silver surfaces of the MTM cavity) so as to vary the refractive index seen by extraordinary waves. 
 
As a further demonstration of mode hybridization, the electric field was calculated using finite element method for the three LC thicknesses under normal incidence in Fig. \ref{SI_EF_LCno} of SI. In Fig. \ref{SI_EF_LCne} of SI the transmittance plots, electric field confinement into the LC cavity ($n_e$), and  electric field distributions along the propagation direction are also shown.

\begin{figure}[!ht]
\includegraphics[width=0.8\columnwidth]{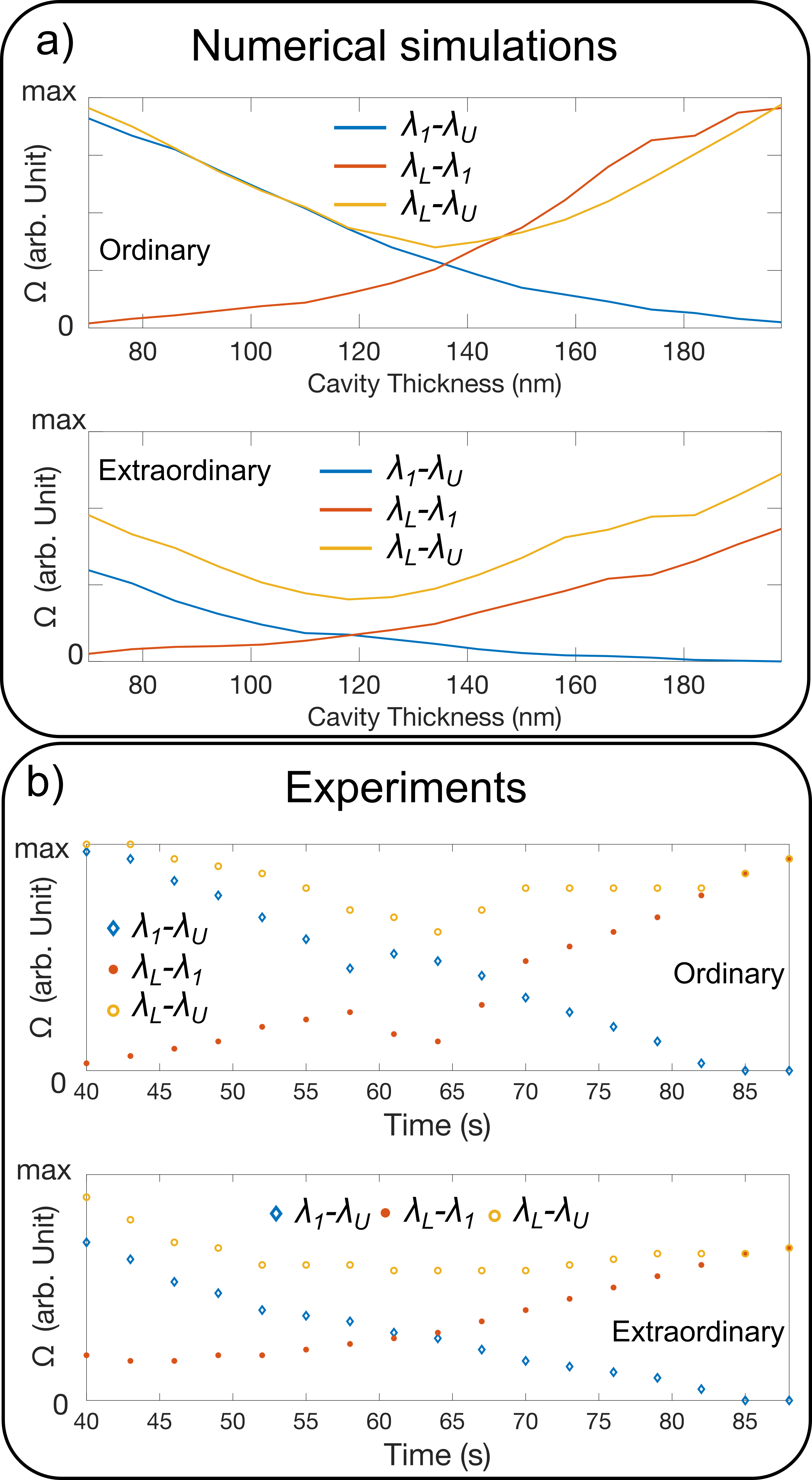}
\caption{a) Numerical and b) experimental distance $2\Omega$ between the wavelengths $\lambda_L$ and $\lambda_U$ for ordinary (top panels) and extraordinary (bottom panels) polarizations calculated for the following resonant mode conditions: $\lambda_U$ approaches $\lambda_1$, $\lambda_L$ departs from $\lambda_1$ and $\lambda_U$ approaches $\lambda_L$ (anti-crossing).}
\label{Sim_ne}
\end{figure} 
In Fig. \ref{Sim_ne}a, the wavelength splitting $2\Omega$ related to the difference $\lambda_L-\lambda_U$, $\lambda_1 -\lambda_U$ and $\lambda_1-\lambda_L$  is shown as a function of the LC thickness. The top and bottom panels show the resonance wavelengths retrieved via a multiple Gaussian fit on the transmittance curves, for the anti-crossing behaviour represented by $\lambda_U$ approaching $\lambda_L$, $\lambda_U$ approaching  $\lambda_1$, and $\lambda_1$ approaching $\lambda_L$, for the ordinary and extraordinary LC refractive index, respectively. The minimum value of $2\Omega$ corresponds to the avoided-crossing point and maximum coupling between same-parity modes. The same behaviour is observed in experiments (Fig. \ref{Sim_ne}b). The slight difference related to the amplitude ($2\Omega$) reported in both plots, numerical and experimental ones, for extraordinary polarization is a consequence of the non negligible effect of changing the surrounding medium around the two MIM resonators.   In SI, we also simulated the angular dependence by varying the incident angle $\theta_i$ from $0^\circ$ to $80^\circ$ in steps of $2^\circ$ for both ordinary and extraordinary polarization (Fig. \ref{SI_angle}. The results show that the three-cavity resonator is not significantly perturbed by the variation of the incident angle, especially for the extraordinary polarization \textcolor{red}.

\section{Conclusions}
In conclusion, we presented a detailed study on how to design and actively tune strongly confined hybrid modes in one-dimensional layered structures working in visible wavelengths. The active control is enabled by a high-birefringence LC in combination with an SFA that can vary the cavity thickness rapidly, continuously, and accurately from several $\mu$m down to direct contact between its metal mirror surfaces. Importantly, we studied numerically and experimentally how the system performs in terms of weak and strong light coupling conditions when an LC film is confined between two MIM cavities. This result has significant practical implications for the development of innovative devices as it enables the possibility to excite multiple resonant modes across the LC cavity. This is of fundamental importance for developing active and reconfigurable devices that can find applications as a platform for optical beam steering devices.. Thanks to the tunability of these photonic modes, the proposed system can be of extremely high importance for bio-sensing where it is necessary to involve high energy modes (short wavelengths, from \SI{450}{\nano \meter} to \SI{530}{\nano \meter} ) excitable in free-space.
Although the cavity resonances were obtained under normal incidence, it is expected that plasmonic modes can also be excited in a multi-cavity metamaterials under oblique incidence, notably without using any coupler (i.e, a grating) to generate evanescent waves. 
To this end, the SFA could be used to study the generation, coupling, and transmission of plasmonic modes in multi-cavity metamaterials as a function of the thickness and refractive index of LC loaded cavity.

\section*{Materials and Method}
\textbf{Samples Fabrication:}\\
The MIM cavities were fabricated by DC/RF sputtering (model Kenosistec KC300C), and they were constituted of Ag and transparent Zinc-oxide (ZnO), respectively with target thickness of 30 and 95 nm, on cylindrical glass lenses. The lenses had diameter of R = 2 cm , the thickness of 4 mm, 60/ 40 scratch/dig surface quality, centration wedge angle $<$5 arcmin, and irregularity (interferometer fringes) $\lambda$/2 at a wavelength of 630 nm. Ag was chosen for its large extinction coefficient $k$ $>$ 1 $\gg$ $n$, ensuring a high reflectivity and an approximately real negative permittivity in the metal layers while ZnO was chosen for its transparency ($n$ $>$ 1 $\gg$ $k$).
For the deposition, the following parameters were used: vacuum $\SI{7}{\cdot 10^{-6}}$, DC power $\SI{100}{\watt}$ for $\SI{62}{\sec}$ for Ag layer while ZnO were deposited using the RF cathode at a power of $\SI{80}{\watt}$ and time of $\SI{31}{\min}$ $\SI{36}{\sec}$.
In order to align the LC layer, a solution photo-active poly(amide imide), denominated JK158  in N-methylpyrrolidone (1 wt.$\%$) was spin-coated on top of the exposed Ag layers. The poly(amide imide) was described in \cite{konieczkowska2015large}. JK158 contains randomly aligned azo-dye molecules that reorient perpendicularly to the polarization direction of UV light to minimise the absorption cross-section. The LCs in contact with the aligned JK158 molecules acquire the same alignment.\\
\textbf{SFA Experiments:}\\
A surface forces apparatus (SFA) Mark III by Surforce LLC, USA was used in the experiments \cite{Zappone_ACSPhotonics_2021,israelachvili1990adhesion}. One of the MIM-coated cylindrical lenses was fixed on a rigid support, whereas the other one was attached to the free end of a double cantilever spring. The surfaces of the lenses were sufficiently far apart from each other to avoid any mechanical interaction and moved freely in contact with a 50 $\mu$L droplet of LC. The droplet was infiltrated between the MIM-coated lenses by capillarity.\\
Transmission spectra were obtained by illuminating the MIMTMIM cavity (consisting of the metal- insulator- metal layers on the lens surfaces and the LC solution) under normal incidence with white light from a halogen lamp. The transmitted light was collected through the entrance slit of an imaging spectrograph (PI Acton Spectra Pro 2300i) aligned with one of the cylindrical lenses, and recorded with a high-sensitivity CCD camera (Andor Newton DU940P-FI). Only a small region of the surface surrounding the contact position was probed, such that $r \leq 0.15 mm \ll R$, equivalent to a sphere-plane geometry \cite{israelachvili2011intermolecular}. A CCD camera image recorded the transmitted intensity I as a function of the wavelength $\lambda$ and position r. Multi-beam interference created resonance peaks in a spectrogram, i.e., local maxima of the 2d intensity function $I(\lambda, r)$, corresponding to constructive interference. Spectrograms were obtained by combining multiple CCD images taken in different but overlapping spectral intervals. Each image was recorded within in less than one second, whereas a spectrogram was completed in less than 20 s.\\
\textbf{Numerical simulations:}\\
The transfer matrix method (TMM) analysis were performed using a script implemented in commercial software Matlab. It uses as input the refractive indices data, retrieved by ellipsometry for the used materials, the layer thicknesses and it allows calculating the spectrum varying the T cavity thickness.\\ 
Finete Element Method (FEM) simulation were performed using COMSOL Multiphysics with the same scheme reported in \cite{Lio2019cavities}. In order to analyse the electric filed $|E|/E_0$, where $E_0$ has been calculated as $E_0 = \sqrt{(P/w)Z_0}$, here P is the input power ($1\:W/m^2$), $w$ represents the area illuminated by the light beam and Z0 is the impedance, in the MIMTMIM system a 1D cutting line have been used to collect the normalized electric filed as function of the structure size and wavelengths. The cutting line has been chose to cover the entire length of the MIMTMIM system plus extra 20 nm in the glass before and after the structure.\\

\acknowledgments{This research is performed in the framework of the bilateral (Italy-Poland) project: "Active metamaterials based on new generation liquid crystals (LCMETA)" funded by the Italian Ministry of Foreign Affairs and International Cooperation and the Polish National Agency for Academic Exchange NAWA. G.E.L and F.R. thank the FASPEC (Fiber-Based Planar Antennas for Biosensing and Diagnostics) - supported by Tuscany region in the Horizon 2020 framework - and the project "Complex Photonic Systems (DFM.AD005. 317). G.E.L. also thanks the research project "FSE-REACT EU" financed by National Social Fund - National Operative Research Program and Innovation 2014-2020 (DM 1062/2021). A.F and R.C thank the project “DEMETRA – Sviluppo di tecnologie di materiali e di tracciabilità per la sicurezza e la qualità dei cibi” PON ARS01 00401. R.K. and J.P. acknowledge the financial support from the MUT University Grant UGB 22 804 from funds for year 2023.\\}

\section*{Authors contribution}
G.E.L. and A.F. equally contributed to this work and wrote the article. G.E.L and R.C. conceived the main idea in the framework of the research project "LCMETA". G.E.L and A.F. performed numerical simulations and samples fabrication. B.Z. performed the SFA measurements and provided theoretical explanations. E.S-B. synthesized and delivered LC photoaligning materials and supported the work with fundamental technical advice. R.K. tested photo-aligning materials. J.P., R.K., C.P.U., F.R., and R.C. provided fundamental support thanks to their expertise in liquid crystals. F.R. provided his expertise on light coupling behavior in complex media to explain the physics behind this work. All authors revised the paper and accepted its contents.
%
%

\newpage
\newpage
\onecolumngrid
\setcounter{figure}{0}
\setcounter{equation}{0}
\setcounter{page}{1}
\renewcommand{\thefigure}{S\arabic{figure}}
\renewcommand{\theequation}{Ax\arabic{equation}}
\newpage
\section*{\centering \textbf{Supplementary Information}}
\section*{Unlocking optical coupling tunability in epsilon-near-zero metamaterials through liquid crystal nanocavities}
\section*{Surface Force Apparatus (SFA)}
\begin{figure}[!ht]
\includegraphics[width=0.5\columnwidth]{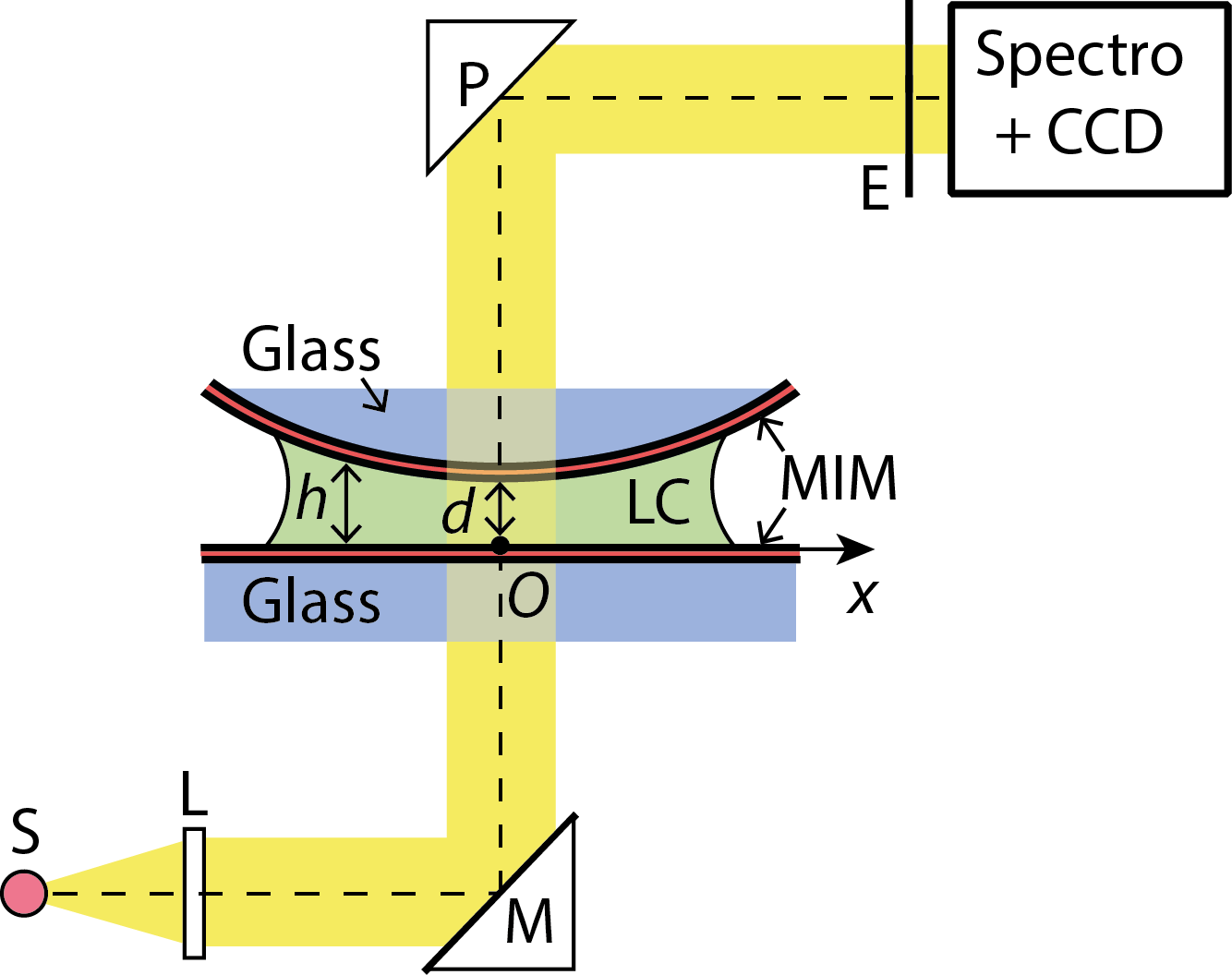}
\caption{Schematic of the SFA setup. White light from a halogen lamp (S) is collimated by a lens (L) and directed by a mirror (M) onto the two MIM-coated cylindrical glass lenses with crossed axes facing each other at distance $d$. The transmitted beam is directed by a right-angle prism (P) into the entrance slit (E) of an imaging spectrograph coupled to a CCD camera. $d$ is the surface distance at the point of closest surface approach, also known as contact point (O). Around this point, the surface distance increases as $h = d + r^2/2R$, where $r$ is the lateral distance from the contact point ($r = 0$) and $R = 2$ cm is the cylinder radius.}
\label{SFA_setup}
\end{figure}

Figure \ref{SFA_setup} shows a schematic of the SFA setup.
\section*{Experimental measurements}
\begin{figure}[!ht]
\includegraphics[width=0.5\columnwidth]{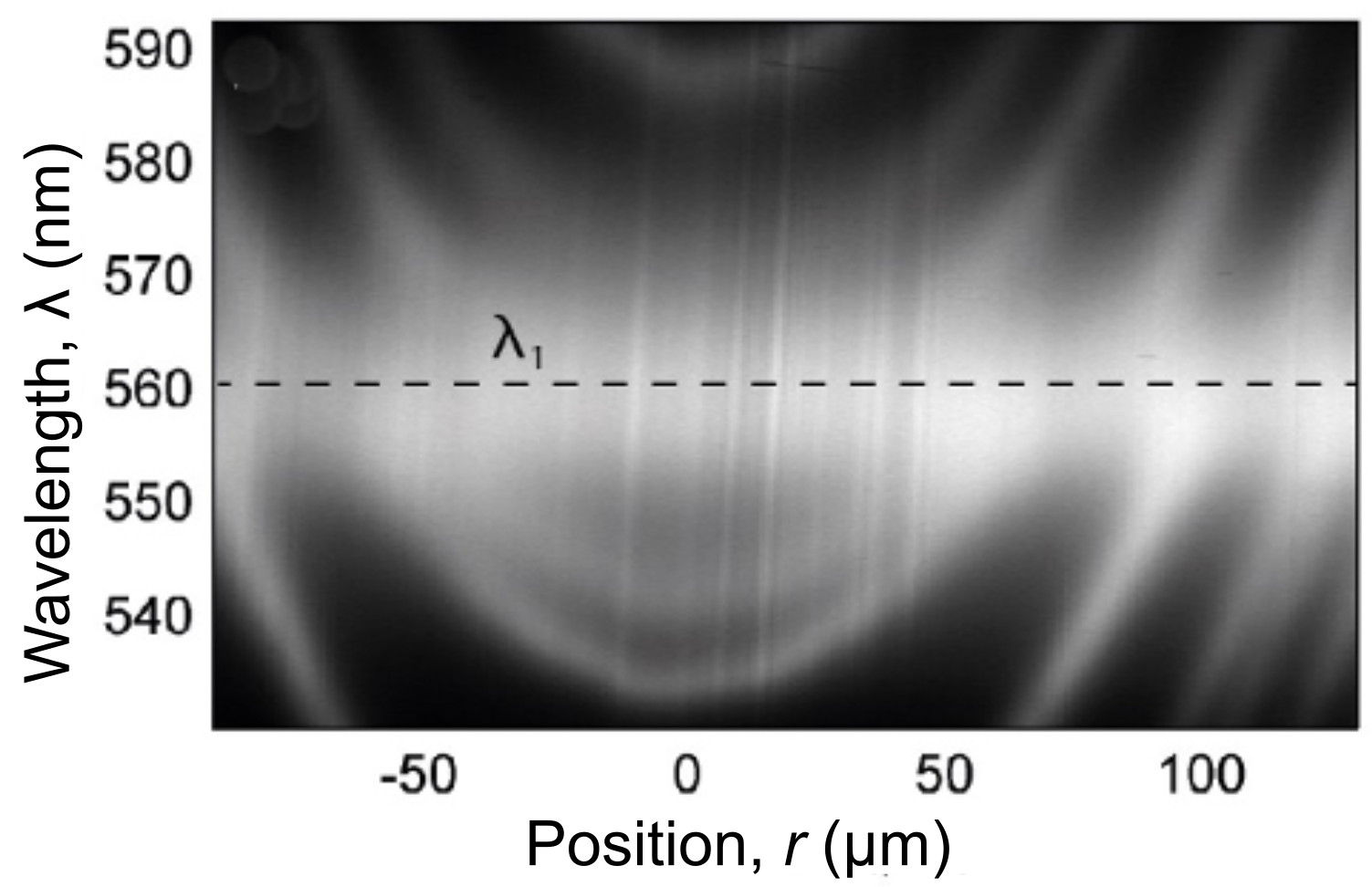}
\caption{Transmitted intensity of the MIMTMIM resonator measured without polarizers.}
\label{SI_exp}
\end{figure}

Figure \ref{SI_exp} shows the transmitted intensity of a MIMTMIM resonator as a function of the wavelength $\lambda$ and lateral distance $r$ from the surface contact position ($r=0$) without polarizers.
\section*{Numerical simulations}
Here are reported the further numerical simulations performed to confirm the trends shown in the main paper. 
\begin{figure}[!ht]
\includegraphics[width=\columnwidth]{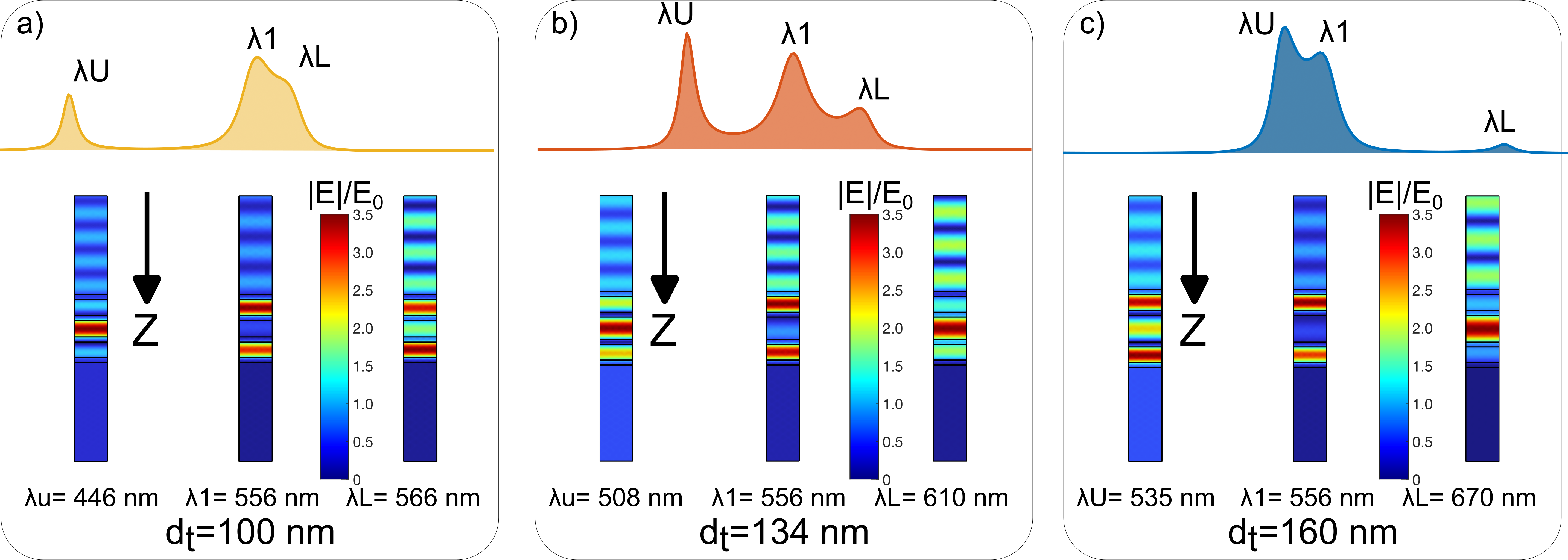}
\caption{Transmitted light plots and normalized electric filed $|E|/E_0$ propagated through the MIMTMIM structure under normal light incidence for the three LC cavities a) $d_T$=\SI{100}{\nano \meter}, b) $d_T$=\SI{134}{\nano \meter}, and c) $d_T$=\SI{160}{\nano \meter} with ordinary refractive index identifying the multiple resonant wavelength $\lambda_U$, $\lambda_1$, and $\lambda_L$.}
\label{SI_EF_LCno}
\end{figure} 
\begin{figure}[!ht]
\includegraphics[width=\columnwidth]{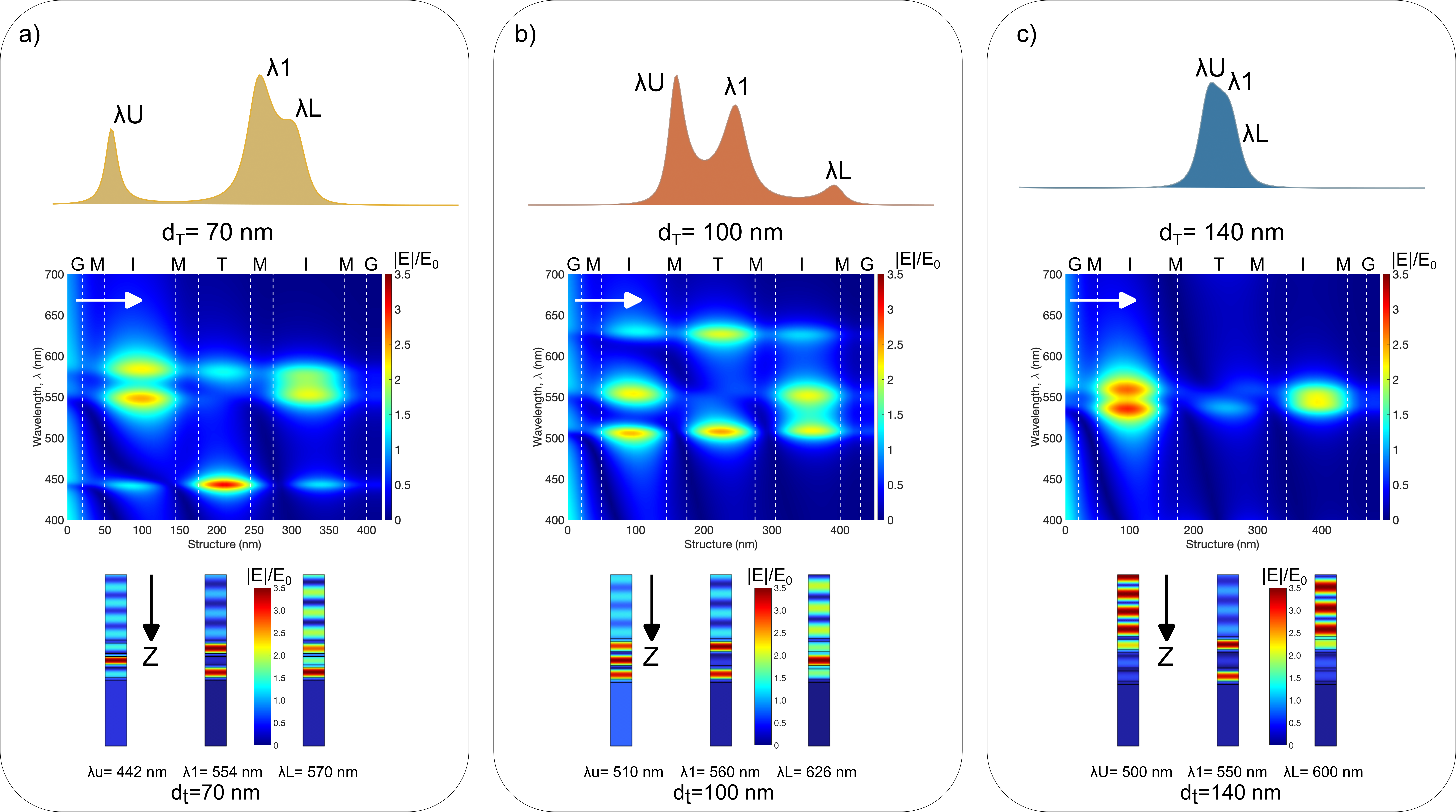}
\caption {Transmitted light plots, electric field maps about the resonant modes and their hybridizations, and normalized electric filed $|E|/E_0$ propagated thought the MIMTMIM structure under normal light incidence for the three LC cavities a) $d_T$=\SI{70}{\nano \meter}, b) $d_T$=\SI{100}{\nano \meter}, and c) $d_T$=\SI{140}{\nano \meter} with extraordinary refractive index identifying the multiple resonant wavelength $\lambda_U$, $\lambda_1$, and $\lambda_L$.}
\label{SI_EF_LCne}
\end{figure}  

\begin{figure}[!ht]
\includegraphics[width=\columnwidth]{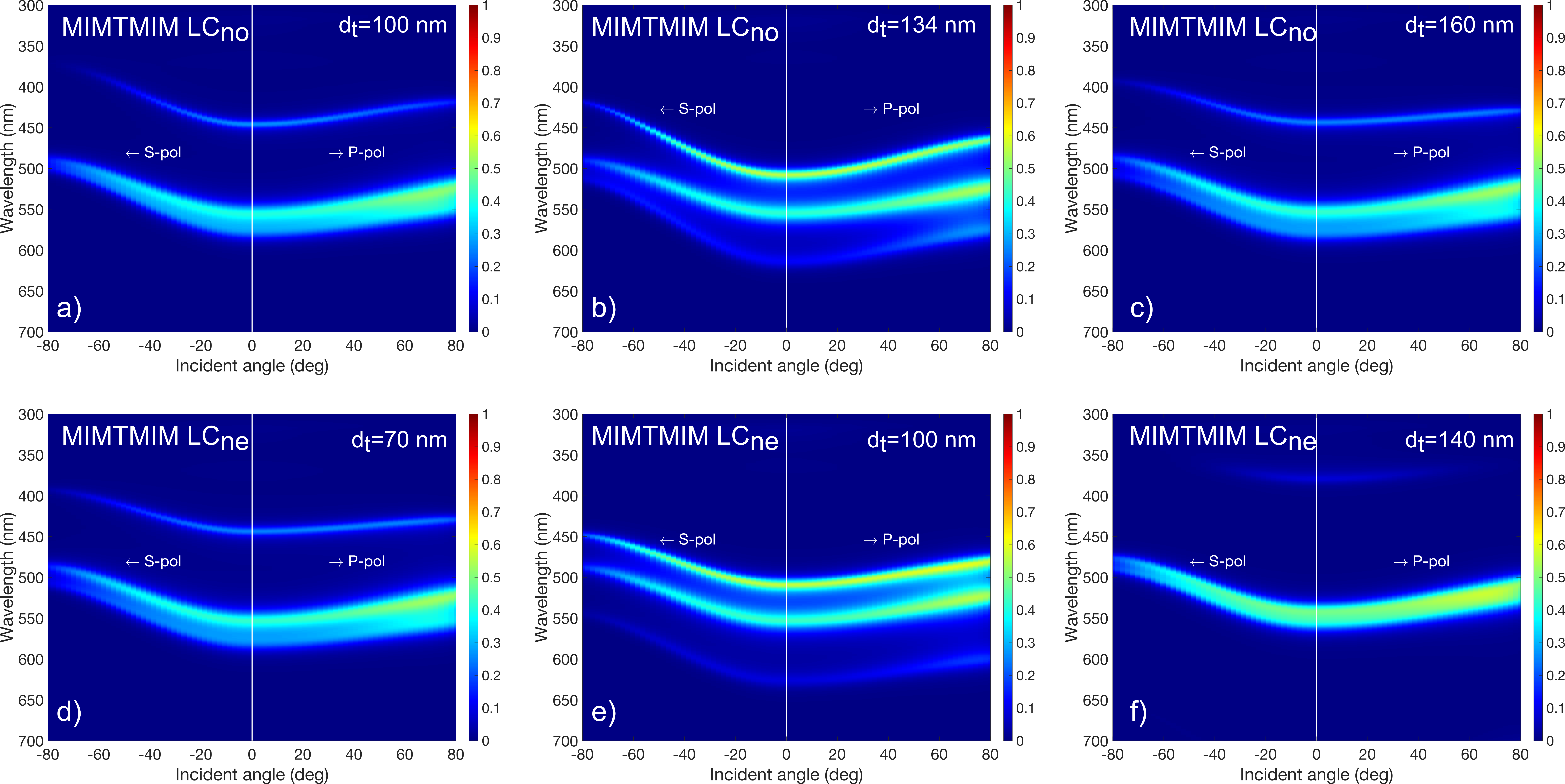}
\caption{Transmitted intensity maps as function of wavelength (from \SI{300}{\nano \meter} to \SI{700}{\nano \meter} by step of \SI{1}{\nano \meter}) and incident angle (varying from $0^\circ$ to $80^\circ$ by step of $1^\circ$). Each map shows on the right side for positive angle the parallel polarization while on the left side (angle that goes from $0^\circ$ to $-80^\circ$) the orthogonal polarization. They are related to MIMTMIM with the LC refractive index equal to $n_o$ and with a cavity thickness of a) $d_t$=\SI{100}{\nano \meter}, b) $d_t$=\SI{134}{\nano \meter}, c) $d_t$=\SI{160}{\nano \meter}, while switching the LC refractive index to the extraordinary one the cavity thickness is d) $d_t$=\SI{70}{\nano \meter}, e) $d_t$=\SI{100}{\nano \meter}, f) $d_t$=\SI{140}{\nano \meter}}
\label{SI_angle}
\end{figure}  
\end{document}